# The Emerging 3GPP 6G RAN Architecture: An Overview

Xingqin Lin

NVIDIA

Email: xingqinl@nvidia.com

***Abstract*—The 6G radio access network (RAN) architecture is emerging as a disciplined evolution of 5G RAN. The 5G baseline introduced modular base station, providing a flexible framework for diverse deployment scenarios and multi-vendor interoperability. The key architectural challenge for 6G RAN is to preserve these benefits while adapting the RAN to new deployment and service requirements. This article reviews the emerging 3GPP 6G RAN architecture with emphasis on boundary selection. It discusses central unit and distributed unit split enhancements, recognition of radio unit as a distinct logical unit, and the RAN-core network interface study, where point-to-point signaling is favored over service-based interface for connectivity services. It also highlights open areas including RAN-core network interface for non-connectivity services, data collection framework, and artificial intelligence for 6G RAN.**

## I. INTRODUCTION

The architecture of the sixth generation (6G) radio access network (RAN) is beginning to take shape, as the 3rd generation partnership project (3GPP) Release-20 6G study progresses [1]. While much public discussion of 6G focuses on radio performance, spectrum, artificial intelligence (AI), and sensing, the long-term success of the system will depend equally on the architecture that surrounds it [2][3]. Architecture determines how functions are placed, how network nodes interact, where interoperability is expected, and where implementation freedom is preserved [4]. These choices shape network deployment, vendor integration, migration strategy, and the ability to introduce new capabilities without creating unnecessary complexity [5]. The immediate architectural baseline for 6G is 5G next-generation RAN (NG-RAN) [6]. The 5G node B (gNB) introduced a modular structure based on the separation of central unit (CU) and distributed unit (DU) functions. It also introduced a further separation between control plane (CP) and user plane (UP) functions inside the CU. NG-RAN provided a flexible framework for diverse deployment scenarios and multi-vendor interoperability. The emerging 6G RAN architecture is expected to refine the NG-RAN framework where the existing design is insufficient and strengthen it where 5G experience has revealed areas for improvement [1].

A central theme of the 6G RAN architecture study is boundary selection. Some boundaries are system-level interoperability points and therefore need to be specified in 3GPP. The CU-DU interface, the CU-CP/CU-UP interface, the RAN-core network (CN) interface, and the interface between 6G node B's (i.e., aNBs) belong to this category [7]. They define how independently implemented functions can interoperate and how the RAN behaves as part of the end-to-end system. Other boundaries are closer to product implementation and deployment specific optimization. These boundaries still matter architecturally, yet they are not always suitable for 3GPP protocol specification. The treatment of the radio unit (RU) illustrates this distinction. In the 5G gNB architecture, the standardized logical decomposition stops at the DU. RU and fronthaul are not visible in the 3GPP logical architecture. In 6G, the RU will be recognized as a distinct logical unit inside the aNB [8]. The recognition is intentionally scoped: the RU functions and the interface between the RU and the rest of the aNB remain outside 3GPP specification. As a result, open RAN (O-RAN) fronthaul, enhanced common public radio interface (eCPRI) based solutions, and proprietary fronthaul can coexist under the same 6G RAN architecture.

The CU-DU split represents a different kind of boundary. Unlike the RU fronthaul, the CU-DU interface is expected to be a 3GPP-defined interface in 6G RAN [7]. Its value comes from pooling, flexible placement, and multi-vendor interoperability [9]. At the same time, the split introduces real technical challenges, e.g., user equipment (UE) context is distributed across entities, and configuration procedures require coordination. For 6G, the goal is to preserve the split deployment value while reducing the coordination cost observed in 5G. The RAN-CN interface is another major architectural decision. For connectivity services, the emerging direction favors a point-to-point (P2P) application model between the aNB and a 6G CN entity, instead of using a service-based interface (SBI) where the aNB and the 6G CN entity interact through services exposed by either side [1]. The comparison with an SBI remains important. The 5G CN has shown the value of service-based architecture for modular CN functions [10]. Some future RAN-CN interactions may also benefit from service-style mechanisms. This may be relevant for non-connectivity services, such as sensing, AI, and service exposure [11]. The key architectural question is how to support such functions without creating a fragmented architecture with different mechanisms for every new service.

This article provides an overview of the emerging 3GPP 6G RAN architecture. We first review the 5G RAN architecture as the baseline for 6G. Then we analyze the rationale, tradeoffs, and potential improvement areas for the CU-DU split in 6G RAN. After that, we discuss the RU as a newly recognized logical unit and explain why RU fronthaul remains outside 3GPP scope. We also examine the P2P and SBI alternatives for the RAN-CN interface. Finally, we conclude the article by pointing out open areas, including RAN-CN interface for non-connectivity services, data collection framework, and AI for 6G RAN.

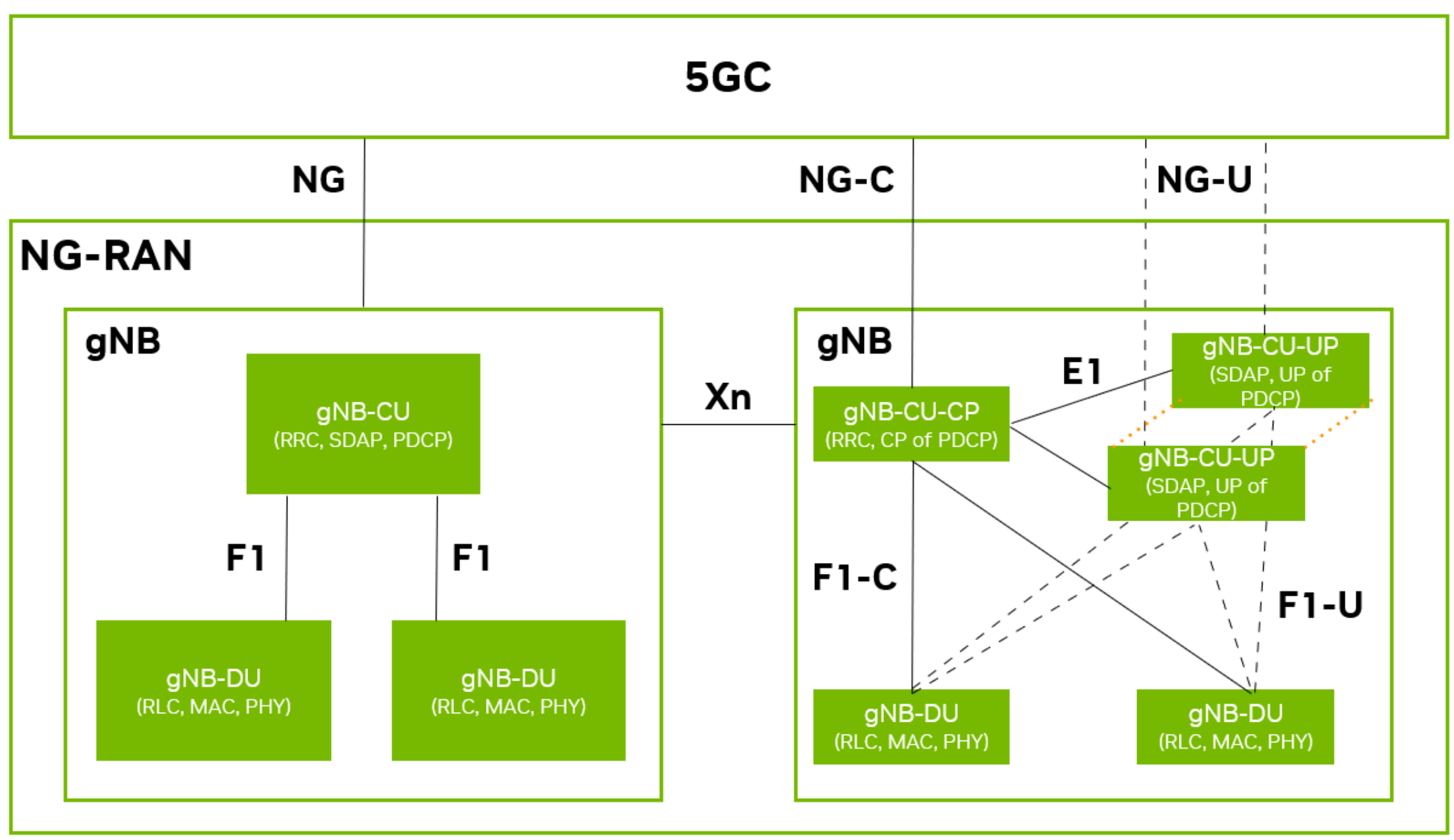

**Figure 1: An illustration of 5G RAN Architecture.**

## II. 5G RAN AS THE BASELINE FOR 6G

The 5G RAN architecture illustrated in Figure 1 provides the immediate baseline from which the 6G RAN architecture is evolving. A clear understanding of the 5G RAN architecture is essential before discussing the emerging 6G RAN. The 5G RAN connected to the 5G CN is called NG-RAN. Its main node is the gNB, which provides new radio (NR) CP and UP protocol terminations toward the UE and connects to the 5G CN through the NG interface. At the highest level, a gNB may be implemented as a single logical node or decomposed into a CU and one or more DUs. The gNB-CU hosts higher-layer functions, while the gNB-DU hosts lower-layer functions. The standardized interface between them is F1. This split allows operators to centralize some baseband functions while keeping time-sensitive functions closer to the radio site. It also allows different deployment models: a compact base station may implement CU and DU together, while a cloud RAN deployment may centralize CUs and deploy DUs closer to radio sites.

The functional split between CU and DU is one of the defining choices in 5G RAN architecture. In the 3GPP logical model, the gNB-CU hosts radio resource control (RRC), service data adaptation protocol (SDAP), and packet data convergence protocol (PDCP) functions. The gNB-DU hosts radio link control (RLC), medium access control (MAC), and physical layer (PHY) functions. This places the CU-DU boundary between PDCP and RLC. The choice reflects a balance between centralization and real-time control. Functions above the split benefit from pooling, centralized mobility handling, and easier UP anchoring. Functions below the split are closer to scheduling, hybrid automatic repeat request (HARQ) timing, and PHY processing, where latency and tight radio coordination matter. The F1 interface has both CP and UP parts. F1-C carries signaling between the gNB-CU and gNB-DU, using F1AP procedures over stream control transmission protocol (SCTP). F1-U carries UP data, typically using general packet radio service (GPRS) tunneling protocol UP (GTP-U) transport. Through F1-C, the CU can configure DU resources, manage UE context, and coordinate radio bearer setup. Through F1-U, UP packets are carried between CU-side PDCP termination and DU-side lower-layer processing. This interface is central to multi-vendor CU-DU interoperability, although real deployments still require careful alignment of supported features, profiles, and operational procedures.

5G also introduced a further split inside the CU: the separation between gNB-CU-CP and gNB-CU-UP. The gNB-CU-CP hosts CP functions, including RRC and the CP part of PDCP. The gNB-CU-UP hosts UP functions, including SDAP and the UP part of PDCP. The interface between them is E1. This separation allows the CP and UP to scale independently. A CP anchor can manage UE context and mobility while UP functions are deployed according to traffic load or local breakout needs. The E1 interface is important because it separates two different scaling problems. CP signaling is driven by mobility, session control, UE state transitions, and radio bearer management. UP processing is driven by throughput, traffic locality, and application placement. In dense or high-capacity networks, UP functions may need to scale faster or be placed differently from CP functions. E1 allows this flexibility while preserving coordinated bearer management and UE context handling.

The gNB connects to the 5G CN through the NG interface. The CP side, often associated with N2, connects the gNB to the access and mobility management function (AMF). The UP side, associated with N3, connects the gNB to the user plane function (UPF). NG-C carries signaling for access, mobility, paging, UE context, and session-related coordination. NG-U carries UP packets using GTP-U. This separation between CP and UP connectivity toward the CN aligns with the broader 5G system architecture, where the AMF handles access and mobility control and the UPF handles UP forwarding. The Xn interface connects NG-RAN nodes. It supports inter-node mobility, context transfer, data forwarding, and load-related coordination.

The 5G architecture therefore contains several standardized boundaries with distinct purposes. NG separates RAN from core. Xn connects RAN nodes. F1 separates CU and DU. E1 separates CU-CP and CU-UP functions. These interfaces made 5G more modular than long-term evolution (LTE) and created a basis for cloud RAN and multi-vendor deployment. They also established an architectural baseline that carries into 6G.

## III. CU-DU Split in 6G RAN

The CU-DU split is one of the most important architecture decisions for 6G RAN. For 6G RAN, the essential question is how much of the 5G model should be retained, how much should be improved, and how the split can support 6G requirements.

### *A. Rationale and Tradeoffs Behind Supporting CU-DU Split*

After extensive discussion, 3GPP agreed that, for 6G RAN, an interface for the CU-DU split shall be supported, and an interface for CU-CP and CU-UP separation shall also be supported, as illustrated in Figure 2. The main reason to support CU-DU split in 6G is the deployment value demonstrated in 5G. Higher-layer functions can be centralized across multiple DUs, creating a shared processing pool. In a monolithic deployment, processing resources are often dimensioned site by site, with each node engineered for local peak demand. This leads to stranded capacity when load varies across cells. With a CU-DU split architecture, traffic and signaling fluctuations from many cells can be aggregated. The aggregate workload is smoother than the workload seen by each individual cell, enabling statistical multiplexing and reducing overprovisioning.

The split also supports more natural scaling. DU resources are driven by radio-facing characteristics such as bandwidth, carrier count, multiple-input multiple-output (MIMO) configuration, and cell coverage. CU resources are driven by higher-layer traffic, UE context, signaling, and bearer processing. When CU-CP and CU-UP separation is also used, the scaling model becomes more granular. CU-CP capacity follows CP demand, while CU-UP capacity follows throughput and UP processing demand. This is attractive for 6G because traffic demand and signaling demand will vary across deployment types. A dense urban network, a private industrial deployment, and a rural macro layer will have different mixtures of coverage, throughput, and mobility requirements. A single monolithic scaling unit may not be satisfactorily matched to that diversity.

The split also provides an interoperability boundary. Without a standardized CU-DU interface, disaggregation becomes an internal vendor design choice. A vendor may modularize its software, yet the operator has no standardized interface at which to mix components from different suppliers. A 3GPP-defined CU-DU interface allows operators to potentially consider CU and DU sourcing separately, introduce vendors incrementally, and avoid making every deployment choice dependent on one supplier's full RAN stack. This is relevant for multi-vendor interoperability in the 6G era. The CU-DU split also complements cloud RAN. A centralized CU can be deployed in an edge cloud, a regional data center, or a central site, while the DU remains closer to the radio. This placement flexibility supports different latency and transport constraints. It may also simplify software lifecycle management. Higher-layer software can be upgraded, monitored, and restored in fewer locations. Redundancy can be implemented through cloud mechanisms. DU failures can remain localized, while the CU maintains broader context and control.

Despite the benefits, there are also technical concerns about the CU-DU split. One concern is UE context handling across the split. In the 5G model, UE configuration and state are shared between CU and DU. The CU controls higher-layer context and generates part of the RRC configuration. The DU controls lower-layer configuration and scheduling-related behavior. Many UE procedures require both entities to maintain a consistent view of the UE. This creates extra signaling, latency, and failure-handling complexity. The issue becomes more critical for bursty traffic and fast state transitions. A UE may resume, send a small amount of data, and become inactive again before a complex CU-DU negotiation has delivered benefit.

Interoperability is another difficult area. A standardized F1-like interface may enable multi-vendor operation at the protocol level, yet optimal performance often depends on implementation details that standards do not capture. Scheduling policy, admission control, mobility robustness behavior, and load balancing strategies are much implementation specific. Capability exchange can help, although it cannot describe every dynamic behavior. The central tension of CU-DU interoperability is that the interface must expose enough information to support robust operation, while avoiding the standardization of vendor algorithms. The UP side has its own challenges. With PDCP placed in the CU and RLC/MAC placed in the DU, F1-U must support flow control, buffering coordination, quality-of-service (QoS) handling, and latency-sensitive forwarding. If the CU pushes data too aggressively, DU buffers can build up and delay packets. If feedback is too conservative, radio resources may be underutilized.

### *B. Potential Enhancement Areas for the 6G CU-DU Split*

The main objective for 6G is to preserve the value of CU-DU split while reducing the coordination cost observed in 5G. This requires improvements in UE context ownership, capability exchange, UP feedback, and interoperability profiling.

UE context ownership is an important improvement area. A split architecture works best when each entity has clear responsibility for the state it controls. Ambiguous ownership leads to repeated negotiation, slow procedure execution, and complex rollback behavior. For 6G, time-critical procedures should minimize CU-DU round trips. This is especially important for connection resume, fast transition to connected mode, bearer modification, mobility preparation, and small-data activity. Where joint CU-DU action is unavoidable, the interface should support compact transactions with clear success, failure, and recovery procedures. The goal is to avoid split-induced latency becoming visible in UE experience. The configuration model also needs careful design. In 5G, the CU and DU both contribute to UE configuration. This can make RRC generation and lower-layer configuration coordination complex. For 6G, the interface should reduce unnecessary

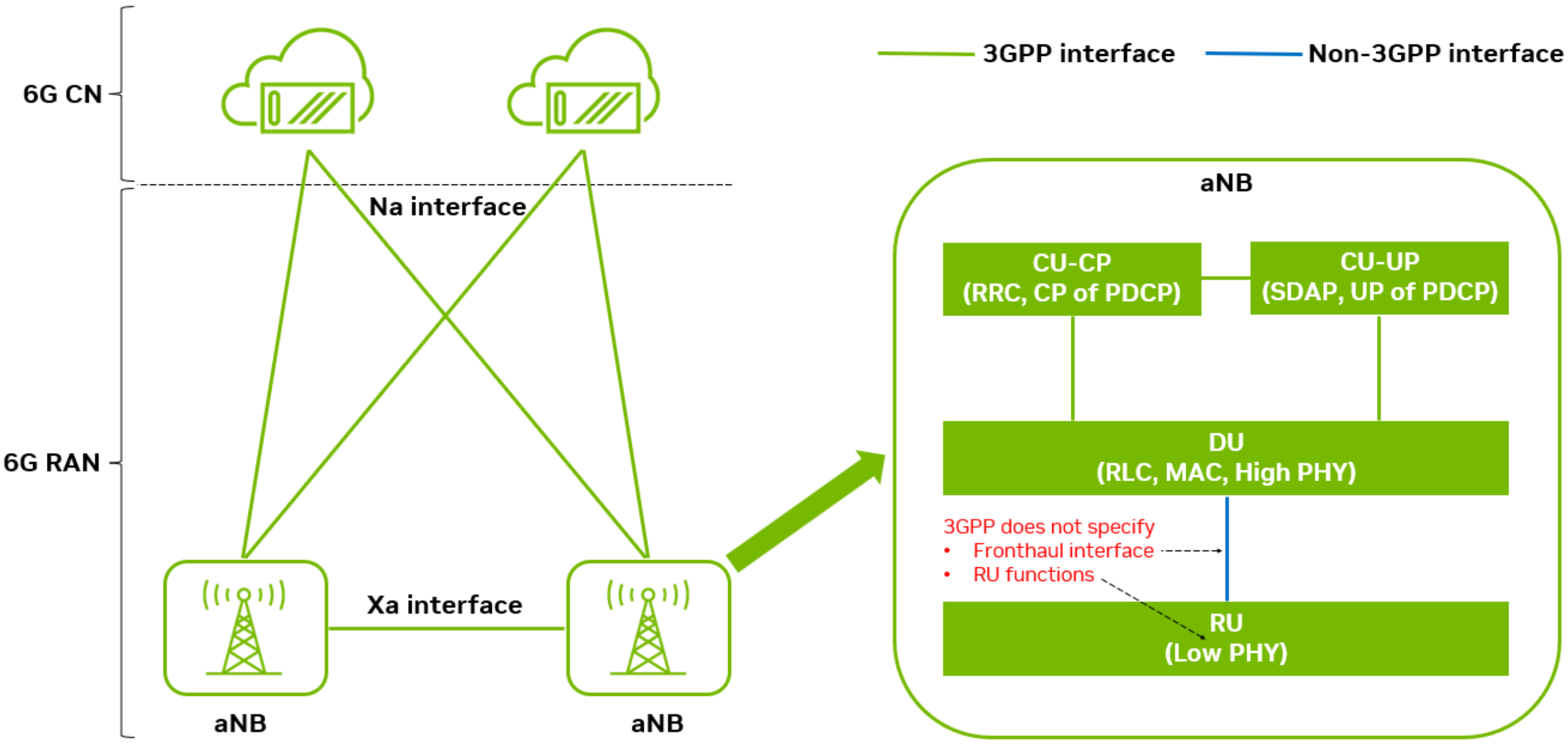

**Figure 2: An illustration of 6G RAN Architecture.**

dependencies between CU-managed and DU-managed configuration.

Capability exchange should also be improved. A CU needs enough information about DU capabilities to make safe and efficient decisions. The DU needs enough information about CU intent to configure lower layers correctly. The challenge is to avoid over-specification. A very detailed capability model may become too large to implement consistently. A very abstract model may leave too much behavior to vendor-specific assumptions. A better 6G approach would define a stable set of mandatory baseline capabilities, a limited set of well-scoped optional capabilities, and testable behavior for common procedures. This would make multi-vendor interoperability more predictable without standardizing proprietary algorithms.

The UP side deserves stronger treatment in 6G. Flow control should become more service aware. The CU should receive useful information about DU buffer status, radio congestion, latency risk, and transmission opportunity without exposing unnecessary scheduler internals. The DU should receive UP data in a way that supports QoS differentiation and avoids buffer buildup. For low-latency traffic, feedback needs to be timely enough to prevent avoidable queuing. For high-throughput traffic, the interface needs to sustain efficient forwarding without excessive control overhead. These requirements suggest that a 6G F1-U like interface should treat latency reporting, congestion indication, and QoS-aware backpressure as core functions.

Interoperability profiling is another potential improvement area. A standardized interface is necessary for multi-vendor operation, but it is rarely sufficient by itself. Commercial interoperability depends on supported feature sets, optional procedure handling, error behavior, and management alignment. 6G should consider clearer baseline profiles for common deployments. Such profiles can identify the minimum behavior needed for robust CU-DU interworking while allowing advanced features to remain optional. This approach would reduce integration burden without restricting vendor innovation.

In short, the 6G CU-DU split should be efficient. The interface should expose the information needed for robust operation, while avoiding the standardization of implementation algorithms. If 6G achieves that balance, the CU-DU split may provide a stronger foundation than its 5G predecessor for 6G RAN deployments.

## IV. RU IN 6G RAN

The recognition of the RU as a distinct logical unit is one of the most visible changes in the emerging 6G RAN architecture, as illustrated in Figure 2. In the 5G RAN architecture, the standardized logical decomposition stops at the gNB-DU. The DU hosts lower-layer functions, while the RU remains an implementation matter. Commercial 5G deployments may use remote radio heads, active antenna units, O-RAN fronthaul, eCPRI, or proprietary fronthaul, yet these elements are not visible in the 3GPP-defined gNB architecture. The 6G architecture changes this abstraction by including a distinct logical RU inside the aNB, giving the aNB a more realistic representation of how modern RANs are deployed. However, the functions hosted by the RU remain outside 3GPP specification scope. The interface between the RU and the rest of the aNB also remains outside 3GPP specification scope. This means the lower-layer split associated with RU-baseband connectivity is acknowledged architecturally, while its technical realization is left to other specifications or implementation choices.

RU has become too important to remain hidden in the base station model. It affects antenna integration, timing, synchronization, power consumption, transport dimensioning, and beamforming implementation. In massive MIMO and higher-frequency deployments, the RU may contain significant radio-side processing and calibration functions. In cloud RAN deployments, many RUs may be distributed across sites while baseband processing is pooled elsewhere. In Open RAN deployments, the RU is also an interoperability boundary [12]. At the same time, specifying the RU functions or fronthaul in 3GPP would require choosing among technical tradeoffs that vary significantly by deployment. A simpler RU can reduce radio-side cost, power, and complexity, although it generally increases fronthaul bandwidth and timing sensitivity. A more capable RU can reduce fronthaul load and support more local radio processing, although it increases RU complexity. Figure 3 provides an illustration of the O-RAN 7.2x split, showing how PHY functions may be distributed across DU and RU [13]. However, macro sites, dense indoor deployments, private networks, and cloud RAN systems may prefer different points in this trade space. For example, to address one limitation of the 7.2x split for uplink massive MIMO, O-RAN has introduced uplink performance improvement (ULPI) techniques, moving additional uplink processing from the DU into the RU. In contrast, the 3GPP 6G RAN architecture recognizes the RU without specifying specific lower-layer split.

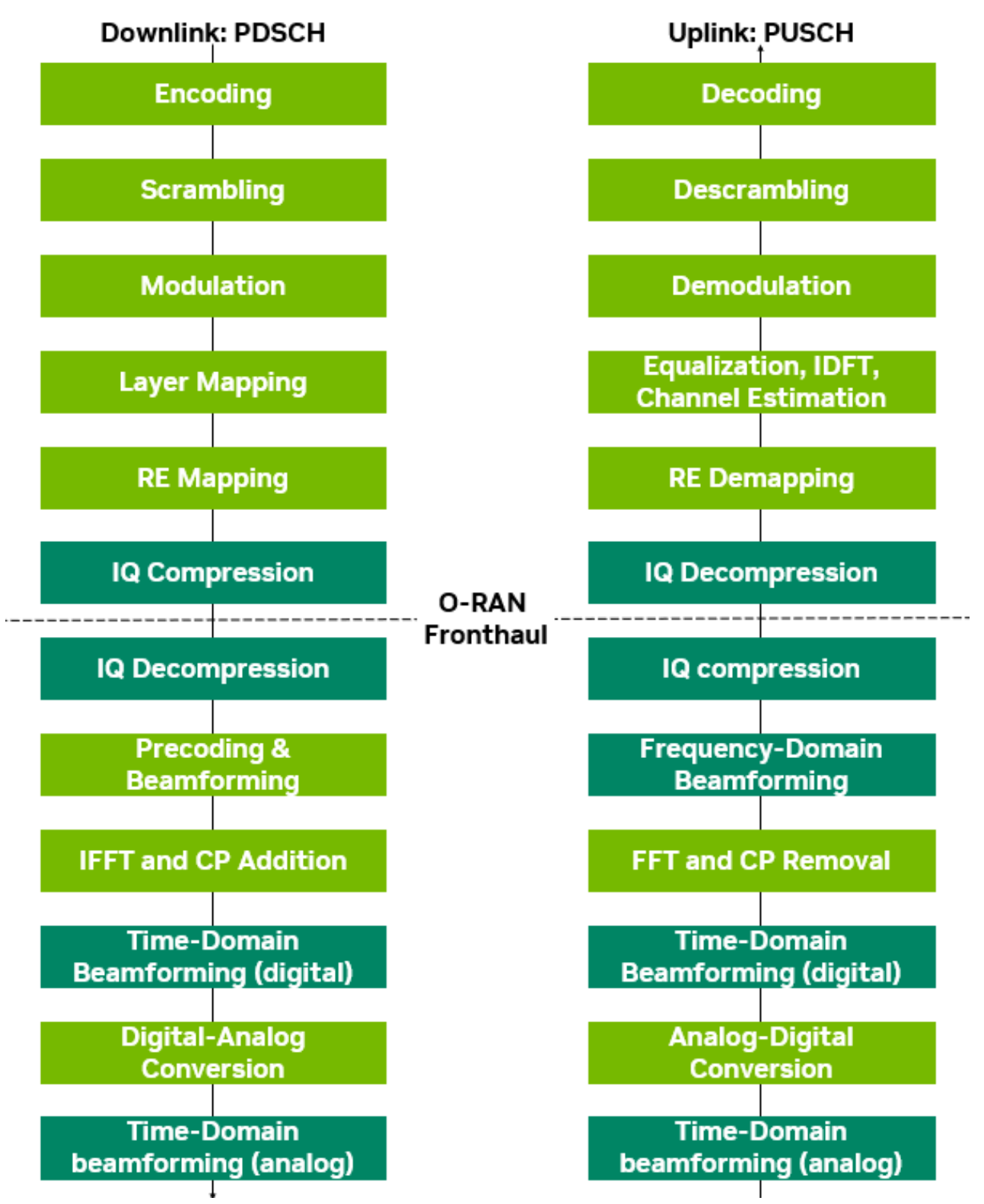

**Figure 3: An illustration of O-RAN 7.2x split. Functions in dark green blocks are optional.**

This scoping also respects the existing standards ecosystem. O-RAN Alliance specifications already provide an open fronthaul framework for deployments that require multi-vendor RU-baseband interoperability. Other deployments may use eCPRI-based solutions or proprietary fronthaul. The aNB architecture accommodates all of these approaches. This avoids duplication of O-RAN work and preserves room for optimized single-vendor and integrated implementations. The RU decision therefore creates a layered interoperability model. Above the RU boundary, the aNB exposes 3GPP-standardized interfaces for interaction with the core network, neighboring RAN nodes, and internal split functions such as CU-DU and CU-CP/CU-UP. At the RU boundary, the implementation can follow O-RAN, eCPRI-based, or proprietary realization. A deployment can be compliant with the 6G RAN architecture while using different fronthaul solutions inside the aNB. This separation allows 3GPP-defined openness where system interoperability requires it, and implementation flexibility where radio/baseband optimization remains deployment specific.

In short, the resulting 6G aNB architecture acknowledges the RU as a distinct logical component of the base station. It avoids imposing one fronthaul design on all deployments. It gives O-RAN a clear role where open fronthaul is needed. It also preserves integrated and proprietary implementations where those are preferable. This balance makes the 6G aNB architecture better aligned with real deployment practice while keeping 3GPP focused on the interfaces and procedures that define the 6G RAN system.

## V. RAN-CN INTERFACE IN 6G

The RAN-CN interface is another key architectural choice in the 6G RAN study [14]. The decision determines the interaction model between RAN and CN: whether connectivity control should be organized as procedure-oriented signaling between paired network entities (i.e., P2P), or as service operations exposed and consumed across an SBI.

The RAN-CN interface must support signaling exchange between RAN and core, CP/UP separation, RAN sharing, network slicing, reliable signaling, and a clear functional split between RAN and CN. It also preserves separation between the radio network layer (RNL) and the transport network layer (TNL), allowing the application protocol to evolve independently from the underlying transport. For the CP, the supported functions include UE context management, non-access stratum (NAS) message transport, protocol data unit (PDU) session management, paging, connected mode mobility support, and configuration transfer between RAN nodes via CN, among others. These are strongly stateful functions, closely tied to UE reachability and mobility.

This functional profile explains why the P2P model is attractive for connectivity services. In the P2P option, the RAN-CN interface is organized as application-layer communication between an aNB and a 6G CN entity using elementary procedures. Procedures may be initiated by either side. They may be UE-associated, such as UE context or NAS transport procedures, or non-UE-associated, such as configuration or warning-message procedures. The model also assumes an RNL-level association per aNB-CN node pair. This gives the interface a clear procedural structure, with defined roles, information exchange, and result handling. The technical strength of P2P lies in ownership of state. Connectivity control is dominated by UE-associated context. The RAN and the CN must agree on UE reachability, mobility state, session-resource handling, paging, and failure recovery. A procedure-oriented interface gives each operation a clear initiator, responder, and outcome. This makes error handling, overload behavior, and partial failure recovery easier to specify. It also matches the operational model that has worked in NG-RAN, where the RAN-CN CP behaves as a reliable signaling interface rather than as a collection of loosely coupled services.

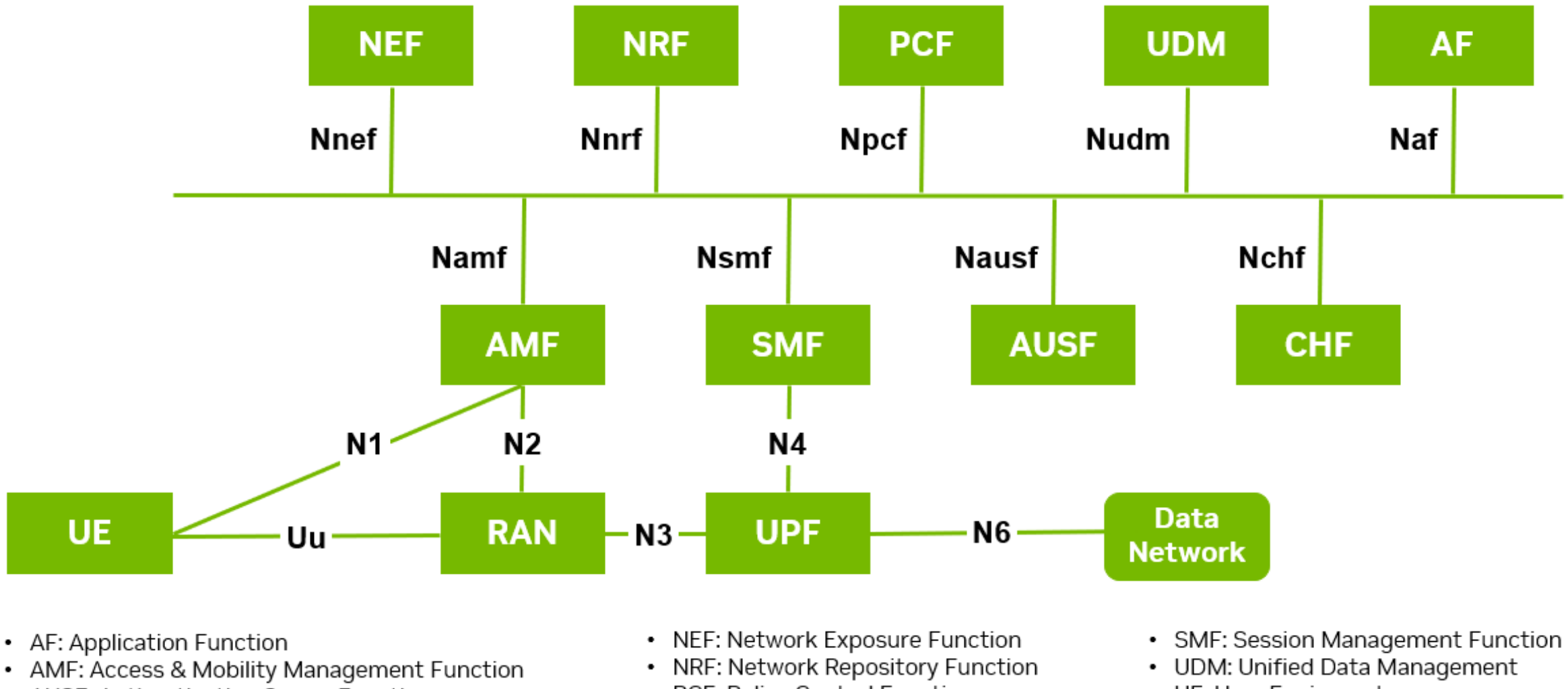

**Figure 4: An illustration of 5G system architecture with P2P RAN-CN interface and SBI interface within 5G CN.**

The P2P application interaction model should be separated from the transport realization. Several transport options are being considered in 3GPP, including SCTP, quick user datagram protocol (UDP) Internet connections (QUIC), and hypertext transfer protocol version 3 (HTTP/3) based realizations. SCTP provides a mature telecom signaling model with message orientation, multi-streaming, and multihoming behavior. QUIC brings integrated transport layer security (TLS) 1.3, faster connection establishment, stream multiplexing, and connection identifiers that decouple the connection from a fixed Internet protocol (IP) address and port. The HTTP/3-based P2P option is interesting because HTTP/3 is treated as part of the transport layer, while RNL messages are carried over long-lived QUIC streams. In that model, HTTP/3 provides a transport substrate for a procedure-oriented RAN-CN application protocol.

The service-based option takes a different view of the RAN-CN boundary. In this model, the aNB and the 6G CN entity interact through services exposed by either side. A service consists of service operations, with one entity acting as producer and the other as consumer. The 3GPP study considers service registration, service update, service de-registration, service exposure, discovery, authentication, and authorization. This framework resembles the service-based principles used inside the 5G CN, where network functions expose services and interact through service operations, as illustrated in Figure 4.

The SBI model may consider two possible communication patterns between the aNB and the 6G CN entity. In the direct communication case, the aNB and the 6G CN entity are configured to communicate with each other directly. The configuration may include information such as service instances and service profiles. In the indirect communication case, the aNB and the 6G CN entity communicate through a communication proxy. The service request includes addressing, discovery, and selection information, allowing the proxy to route the request to a suitable service producer. The proxy may be preconfigured with the necessary information or may perform discovery through a discovery function. Direct SBI can offer a more explicit service abstraction while keeping routing simple. Indirect SBI can support more dynamic service selection, proxy-based routing, and flexible deployment of service producers.

SBI has advantages in domains where service discovery, modularity, and dynamic service composition are central. This is why the 5G CN uses service-based architecture internally. Core functions expose services, discover peers, and interact through service operations. Such a model is powerful when the network contains many service producers and consumers with flexible deployment and selection requirements. The RAN-CN connectivity interface has a different character. The RAN needs a predictable relationship with the core function responsible for access and mobility control. Paging, UE context management, NAS transport, and connected-mode mobility are time-sensitive and stateful. Mapping these procedures into service operations is possible. The issue is whether the additional service framework provides enough benefit for the basic connectivity path. Service registration, discovery, producer/consumer modeling, and authorization add architectural flexibility. They also add operational and standardization complexity to procedures that require deterministic behavior.

This is the main reason the emerging 6G RAN favors P2P for connectivity services. It preserves a direct procedural model for the most fundamental RAN-CN interactions while still allowing modernization of the transport layer. It also keeps the RAN/CN functional split clear. The core remains responsible for access and mobility management, session coordination, and NAS termination toward the UE, while the RAN remains responsible for radio control and mobility execution. A P2P RAN-CN interface provides a stable boundary between these responsibilities. However, service-based mechanisms may still have a role for other kinds of RAN-CN interaction. Sensing, AI, data collection, service exposure, and compute-related support may have different interaction patterns from UE mobility and paging [15]. Some of those functions may benefit from service-style mechanisms, especially when discovery, authorization, or exposure to multiple consumers become important.

## VI. CONCLUSION AND FUTURE OUTLOOK

The emerging 6G RAN architecture reflects a disciplined evolution of 5G RAN. Its main direction is now becoming clear: a standalone 6G RAN, a 6G aNB with a distinct logical RU, support for CU-DU and CU-CP/CU-UP split interfaces, a RAN-domain interface between aNBs, and a RAN-CN connectivity interface based on a P2P application model. The common theme is boundary selection. Interfaces that define system-level behavior are being kept within 3GPP scope and are expected to support multi-vendor interoperability. Boundaries that depend heavily on radio implementation choices, especially RU fronthaul, are left to O-RAN, eCPRI-based, or proprietary realizations.

While significant progress has been made, several important aspects remain open and require further study.

- **RAN-CN interface for non-connectivity services**: While P2P model fits RAN-CN connectivity services, how to handle non-connectivity services (e.g., sensing, AI) remains open. The SBI option studied for RAN-CN, including direct and indirect communication through a proxy, may therefore remain relevant beyond baseline connectivity. The challenge will be to avoid creating fragmented architecture with different mechanisms for every new service.
- **Data collection framework:** 6G RAN is expected to support standardized measurement data collection and transport for existing and new services. This includes data that may be used for AI, sensing, performance monitoring, service awareness, and energy optimization. The architecture must define what data is collected, where it is transported, how it is governed, and how operator control is enforced.
- **AI for 6G RAN:** Training, inference, monitoring, and feedback operate on different time scales and may reside in different parts of the system. A successful architecture should support these workflows without introducing excessive interface complexity. 6G RAN should distinguish training data, inference input, model performance feedback, and operational telemetry, while keeping the number of standardized mechanisms manageable.

The next stage for 3GPP is to complete the Release-20 study and specify 6G RAN architecture in Release 21. A successful 6G RAN architecture should provide both stability and flexibility: stable enough for multi-vendor interoperability at standardized boundaries, and flexible enough to support diverse implementations and deployments, existing and new services, and AI-native network operation.